\documentclass{article}

\usepackage{arxiv}
\usepackage{caption}
\usepackage{graphicx}
\usepackage[utf8]{inputenc} 
\usepackage[T1]{fontenc}    
\usepackage{hyperref}       
\usepackage{url}            
\usepackage{booktabs}       
\usepackage{amsfonts}       
\usepackage{nicefrac}       
\usepackage{microtype}      
\usepackage{lipsum}
\usepackage{listings}

\title{The transformation process from in-campus classes into online classes due to COVID-19 pandemic situation - the case of Higher Education Institutions in Kosovo}

\author{
  Ereza Baftiu, Krenare Pireva Nuçi  \\ 
  Department of Computer Science and Engineering\\
  University for Business and Technology\\
  10000 Prishine, Kosovo \\
  \texttt{krenare.pireva@ubt-uni.net} \\

}

\begin{document}
\maketitle    
\begin{abstract}
The COVID-19 pandemic has caused changes in terms of traditional teaching globally. In Kosova context, the Universities have found the transition from teaching in class to online classes quite challenging. This study investigates the transformation process from in-campus classes to online classes from the technical perspective within five Higher Education Institutions (HEI) in Kosovo. The data was collected using the qualitative methods and its analysis followed the 3C Lichtman approach. The results show that each of the Universities followed a different approach, by using either their limited premises infrastructure or using additional cloud infrastructure.

\end{abstract}

\section{Introduction}
\label{sec:introduction}
Due to the COVID-19 pandemic, which burst out this year, an entire education system was forced to make a switch from in-campus classes into online classes. This was a challenging situation for all countries worldwide, especially with developing countries that are less experienced in online classes. Technology has played a greater role with respect to the Education system, which facilitated the process of learning for the acquisition of knowledge, skills, and values \cite{kolar2020experiences}. 
In Kosovo, Higher Education Institutions continued to have online lectures using various video conferencing platforms such as Google Meet, Zoom, BlueJeans, BigBlueButton, to name a few. Video conferencing platforms enable online communication for audio meetings, video meetings, and seminars, with built-in features such as chat, screen sharing, and recording. In the last decades a number of Universities have established their own eLearning platforms \cite{stefanov2020learning} which now played a key role when dealing with the online learning process. However, the unfamiliarity of students and professors with the use of the technology and a considerable number of them having difficulties adapting fast to the current situation created a void situation in the overall education system. In this regard, the students, professors, and management level needed to come up with an ad-hoc plan in order to continue the process of teaching and learning and be able to re-adapt the plan based on the readiness of all stakeholders and their current progress.

\subsection{Aim and Research Question}
The aim of this research is to document the situation of Kosovo Universities with respect to COVID-19 reaction and identify their technology and network infrastructure for providing online classes once the country was experiencing the total lockdown situation. 

This paper aims to contribute in the following points:
\begin{enumerate}
    \item Identify the challenges that Universities experienced to switch from in campus classes into online classes
    \item Analyse the use of network infrastructure and related technology for providing online classes from a number of Universities in Kosovo
\end{enumerate}

\section{Related Work}
\label{sec:relatedWork}

Today, due to the COVID-19 situation, the Universities needed to switch to online classes for 100 percent of their services. This section describes the research that investigated a number of Universities worldwide with respect to this situation and tried to grasp the knowledge in the context of the particular countries, how these institutions have been affected during this pandemic, specifically on demands of IT services and how students and teachers are affected from this situation. 
This section will give the fundamental concepts of tools that are used by Universities with a focus on video conferencing tools as part of real-time online learning, especially the chosen architecture and their topology.

According to \cite{bhakti2020web}, video conferencing as part of real-time online learning allows two or more people to have real-time communication through voice and images. Real-time communication (synchronous communication) can be through audio, video, and chat that would help raise the level of interactivity between the participants. Today, there a number of service providers that offer video conferencing services, however according to an online article \cite{kullberg2018implementing}  we will list only 5 of them, which are also used within the Universities in Kosovo. Generally, there are two categories of providers, (i) those that offer ready-made platforms and the institutions can only pay for their service, and (ii) those providers that offer their services as an open-source platform. The latter offer their code for free, so the institutions can use, adapt or even further develop in the context of their institutions' needs and requirements. 
Table \ref{table:1}, lists a number of commercial providers, but all of them also offer a limited free version. 

\begin{table}[]
\centering
\caption{Commercial video conferencing service providers}
\label{tab:my-table}
\resizebox{\textwidth}{!}{%
\begin{tabular}{lllll}
\hline
Providers   & Meeting Duration & Users & With the same account                                              & Address                                                                                \\
\hline
Zoom        & 40 min           & 100   & Unlimited                                                          & https://zoom.us/                                                                       \\
\hline
WebEx       & 40 min           & 50    & 1 month                                                            & https://webex.com/                                                                     \\
\hline
GoToMeeting & 40 min           & 250   & 14 days                                                            & \begin{tabular}[c]{@{}l@{}}https://gotomeeting.com\\ /en-ie\end{tabular}               \\
\hline
GoogleMeet  & Unlimited        & 250   & G Suite users                                                      & https://meet.google.com/                                                               \\
\hline
BlueJeans   & 2 hours          & 100   & \begin{tabular}[c]{@{}l@{}}Without \\ creating an acc\end{tabular} & \begin{tabular}[c]{@{}l@{}}https://www.bluejeans.com\\ /products/meetings\end{tabular}

\end{tabular}%
}
\label{table:1}
\end{table}

\begin{figure}
 \begin{center}
\includegraphics[width=13cm]{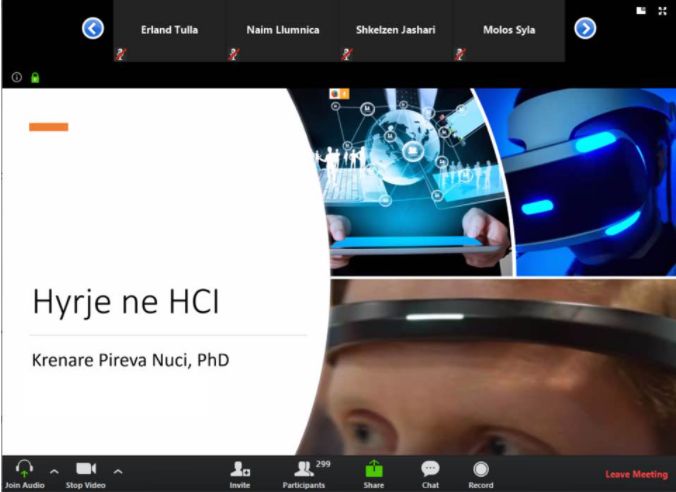}
\caption{Online teaching through Zoom.us.} \label{fig2}
\end{center}
\end{figure}

Through an account (see Table \ref{table:1}), the users can create an unlimited number of video conferences, but their duration is limited to certain minutes (with respect to users of zoom.us is limited when we have 3-100 users, while unlimited for less), in addition to Google Meet which offers unlimited video conferencing duration for up to 250 users, it also demands the Universities to be part of G Suite clients, in Kosovo context, most educational institutions are G Suite clients. Another feature where these providers vary is the number of users in that video conference. Discussing free versions, zoom(see Fig.~\ref{fig2})
and WebEx video conferencing services are open to 100 users, while GoToMeeting and Google Meet for 250 users each.
It is worth noting that with an email account, the free version is limited to one month on WebEx (90 days at the time of the COVID-19 pandemic), and 14 days on GoToMeeting, while other providers have no limit at all \cite{bhakti2020web}.

However, if we think and plan that such services can be provided to a larger number of users and for a longer time, a good opportunity is once deciding on the adaptation or further development of one of the providers listed in Table \ref{tab:table-2} that offer their platforms as open source.

\begin{table}[]
\centering
\caption{Providers of open source platforms for video conferencing services}
\label{tab:my-table}
\resizebox{\textwidth}{!}{%
\begin{tabular}{lllll}
\hline
Providers     & Technology       & Topology    & Users   & Address                     \\
\hline
BigBlueButton & Java             & MCU         & 1-1000+ & https://bigbluebutton.org/  \\
\hline
Licode        & C++              & MCU         & 1-1000+ & https://lynckia.com/licode/ \\
\hline
Jitsi         & Java, JavaScript & SFU         & 1-1000+ & https://jitsi.org/          \\
\hline
Kurento       & C, C++           & MCU         & 1-1000+ & https://kurento.org/        \\
\hline
Jami          & C,C++            & Distributed & 1-1000+ & https://jami.net/ 

\end{tabular}%
}
\label{tab:table-2}
\end{table}

All providers listed in Table \ref{tab:table-2}, offer their free platforms (protected by open source licenses) with the possibility of their adaptation and further development for the institutions.

\begin{figure}
\includegraphics[width=\textwidth]{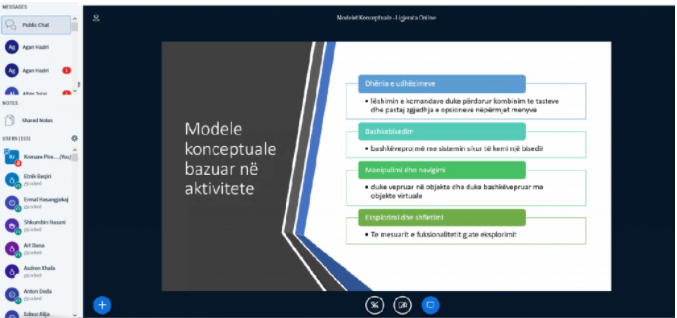}
\caption{Online teaching through the Big Blue Button} \label{BigBlueButtonKNP.png}
\end{figure}

However, to determine for one of these alternatives listed in Table \ref{tab:table-2} it is good to set measurement criteria that would facilitate the decision-making process. For example, for the particular institution if the number of users in real-time is very high then the institution should have knowledge of existing typologies \cite{erezabaftiuthesis}, then with what technology it has been developed, what functions it offers (screen sharing, board writing, handshake by users, providing images through the camera, recording lectures, communication via chat) and so on.
Bigbluebutton \cite{bhakti2020web} is an open-source web conferencing system for online learning. This means you have full access to BigBlueButton’s source code under an open-source license and adapt the features as it is appropriate for your institution. These features include real-time sharing of audio, video, presentation, and screen – along with collaboration tools such as whiteboard, shared notes, polling, and breakout rooms. BigBlueButton can record your sessions for later playback. BigBlueButton has built-in integrations with all the major learning management systems (LMS), including Canvas, Jenzabar, Moodle, Sakai, and Schoology (see Fig.~\ref{BigBlueButtonKNP.png}).

Licode is an open-source project that allows you to include real-time communications like streaming or video conference in your web application in a very easy and fast way.
Jisti \cite{kullberg2018implementing} is a set of open-source projects that allows you to easily build and deploy secure videoconferencing solutions. At the heart of Jitsi are Jitsi Video bridge and Jitsi Meet, which let you have conferences on the internet, while other projects in the community enable other features such as audio, dial-in, recording, and simulcasting. 
Kurento is an open-source software project providing a platform suitable for creating modular applications with advanced real-time communication capabilities. Kurento is a WebRTC media server and a set of client APIS making simple the development of advanced video applications for www and smartphone platforms. Kurento Media Server features include group communications, transcoding, recording, mixing, broadcasting, and routing of audiovisual flows.

Jami (formerly GNU Ring and SFLphone) is an open-source SIP-compatible softphone and SIP-based instant messenger for Linux, Microsoft Windows, Mac OS X, and Android. With this software, you can make calls, create conferences with multiple participants, share media, send text messages during calls or out of calls. Features include an unlimited number of calls, instant messaging, call recording, audio and video calls with multi-party audio, and experimentally video conferencing.

From earlier research, back in 2005, the research investigated the use of learning management systems and other applications in Universities \cite{arkorful2015role}. This research resulted that only two institutions reported not currently using LMS, all other institutions have used at least one LMS (In total participated 111 Institutions). Seven institutions (37 percent) reported using only the institution-wide LMS, and did not use other systems. They all reported significant and ongoing investment in IT networks to support on-campus activity and/or distance learning, and many reported adequate functionality/bandwidth to support eLearning in the short-to-medium term. On-campus, the standard model was Ethernet linked by fiber-optic connections between buildings/campuses (typically one-gigabit backbone, and around 100 megabits to the desktop) –with some institutions reporting plans to upgrade to one gigabit Ethernet within buildings. To give an indication of capacity, a number of institutions reported operation-wide multicast streaming functionality or cited imminent upgrades to this effect. Some institutions reported examples of ongoing dependence on Bayonet Neill-Concelman BNC cables as well as Ethernet. Many institutions were connected to both the commodity Internet and dedicated, higher bandwidth academic networks.

This gives enough argument to claim that universities have started to invest in infrastructure which was a plus once we started to experience the pandemic situation.
According to \cite{adnan2020online} a study was conducted for higher education students towards compulsory digital courses and distance learning in universities due to the Coronavirus pandemic (COVID-19). Undergraduate and postgraduate studies were surveyed to find out their perspectives on online education in Pakistan. The findings highlighted that online learning may not produce the desired results in developing countries like Pakistan, where the vast majority of students are unable to access the internet due to technical and monetary issues. This study \cite{stefanov2020learning} listed several problems that were highlighted by the students, such as: 
\begin{enumerate}
\item	Lack of face-to-face interaction with the instructor, 
\item response time, and 
\item lack of traditional classroom socialization

\end{enumerate}

In contrast to \cite{stefanov2020learning} which conducted the students' perspective toward online learning, the research paper \cite{kebritchi2017issues} investigated the challenges for teaching successful online courses in Higher Education. A review of the literature using the COOPER framework was carried out to identify such issues. The COOPER framework \cite{emrouznejad2010cooper} involves six interrelated phases: (i) \textbf{C}oncepts and objectives, (ii) \textbf{O}n structuring data, (iii) \textbf{O}perational models, (iv) a \textbf{P}erformance comparison model, (v) \textbf{E}valuation and (vi) \textbf{R}esults and deployment. Once approaching online learning, in \cite{bhakti2020web} the three main categories of findings were identified, such as (i) issues related to online students, (ii) instructors, and (iii) content development part. With respect to students, the authors included student expectations, readiness, identity, and participation in online courses. From the teacher's perspective, the authors included the possibilities of changing faculty roles, face-to-face transition online, time management, and teaching styles. And finally, content issues included the role of instructors in content development, the integration of multimedia into content, the role of guiding strategies in content development, and considerations for content development. 
Further, the authors in \cite{krishnamurthy2020future}, investigates how the coronavirus pandemic is forcing global experimentation with distance learning, by emphasizing the shift to virtual learning for the future of higher education institutions. There are many indications that the pandemic crisis will transform many aspects of life including education if distance learning proves to be successful. The paper \cite{ali2013network} emphasizes in order for the Universities to prepare the students to be successful for the future they should provide a strong and flexible learning infrastructure, capable of supporting and providing ubiquitous access to technology tools that allow students to create, design and explore. Essential components of an infrastructure capable of supporting transformative learning experiences include (i) Ubiquitous connectivity, (ii) Powerful learning devices, (iii) High-quality digital learning content, (iv) Responsible use policies (RUPs). 
According to \cite{kolar2020experiences}, the COVID-19 pandemic challenged the education system worldwide and forced teachers to switch to an online overnight teaching mode. Many academic institutions that were previously reluctant to change their traditional pedagogical approach had no choice but to switch entirely to online teaching-learning. The authors shed light on the growth of EdTech startups during times of pandemic and natural disasters and include suggestions for academic institutions on how to address the challenges associated with online learning. However, the authors in paper \cite{meyer2010infrastructure} evaluated the availability and effectiveness of administrative support for online teaching faculty. When administrators make decisions about the infrastructure needs of an actual online teaching program, these decisions are often based on the advice of outside experts. In this context, the online teaching faculty should be the best source of advice and information to see what fits and what does not fit their need. Four main factors were the focus of this study:
\begin{enumerate}
    \item Faculty perceptions of what elements are important for developing a successful online teaching program
    \item Perceptions of which of those elements were successfully implemented in the specific institution
 \item Factors that serve to increase faculty participation in an online curriculum, and what factors hinder their involvement
\item Faculty perceptions of the expected clarity and effectiveness of the Matrix.
\end{enumerate}

Paper \cite{Litchman3C} addresses the possibility of further decision-making for the academic future during any type of disaster. The study emphasized the essential requirements of online teaching-learning in terms of education during the COVID-19 pandemic and show how the resources of educational institutions have effectively transformed formal education into online education through the help of virtual classrooms and other online equipment [10]. The research conducted treated three different aspects: (i) discover the different forms of online teaching-learning methods used during the COVID-19 pandemic, (ii) the perceptions of teachers and students on online teaching-learning during the COVID-19 pandemic, (iii) the challenges faced by teachers and students in adapting to the online teaching-learning process during the COVID-19 pandemic. According to \cite{stefanov2020learning} the whole lecture was organized online due to COVID-19, and keeping the students engaged and more active virtually was challenging. Therefore, the first experiment targeted the engagement and interaction among professors and students in distance learning during the quarantine experience. Initially, the lecture started on a specific topic in HCI using the Zoom and BigBlueButton video conference platforms, and after covering the main learning objective of the specific topic, an in-lecture quiz using online quiz platforms was offered to the students to trigger their presence and keep them more engaged with the lecture. The in-lecture quiz was performed using different tools, where the control group used Google Form Quiz, and the experimental group used Kahoot!. Since the Kahoot! platform has more gamification components, the authors were curious to know whether these components, besides the usability approach, would contribute to increased engagement of the students in the learning process and interactivity among the professor and students.

Lately, the research paper \cite{reimers2020framework} proposed a framework to guide an education on how to respond to the pandemic situations in order to support education decisions to establish and implement effective education. Research shows that education system leaders and organizations develop plans for continuing education through alternative modalities. They claimed that if an online education strategy is not feasible, alternative distribution tools may be developed including television programming, if it is a partnership with television stations is feasible, podcasts, radio broadcasts, and instructional packages either in digital or paper form (same approached followed also by Kosovo Institutions for primary and secondary schools). 
And finally, the article in \cite{stefanov2020learning} investigated the approach of offering online learning from technology and infrastructure in terms of hardware and software. Further, it investigates how much interaction was provided from students in the classroom and how much in the online classes. According to the interaction of students in online classes, the results show that the interaction in the classroom compared to online learning resulted in only 3 percent less. 
Inspired by a number of reviewed articles, we came to a conclusion to continue our current research with respect to the Kosovo Higher Education Institution perspective, and dig deeper to explore the situation how the institutions were able to adapt agilely in this unprecedented situation.

\section{Methodology}
\label{sec:methodology}
\begin{figure}
 \begin{center}
\includegraphics[width=10cm]{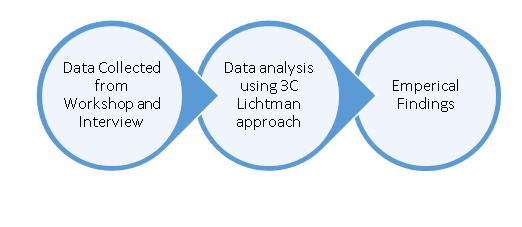}
\caption{Methodology Processes} \label{MethodologyProcesses.png}
\end{center}
\end{figure}

This section gives an overview regarding the methodology used and it describes how the data are collected and analyzed. (Figure ~\ref{MethodologyProcesses.png}), describes the methods used for each phase of the research.

Each of the processes as part of the methodology is described further in the following sections. 

\subsection{Data Collection}
\begin{figure}[h!]
 \begin{center}
\includegraphics[width=150mm]{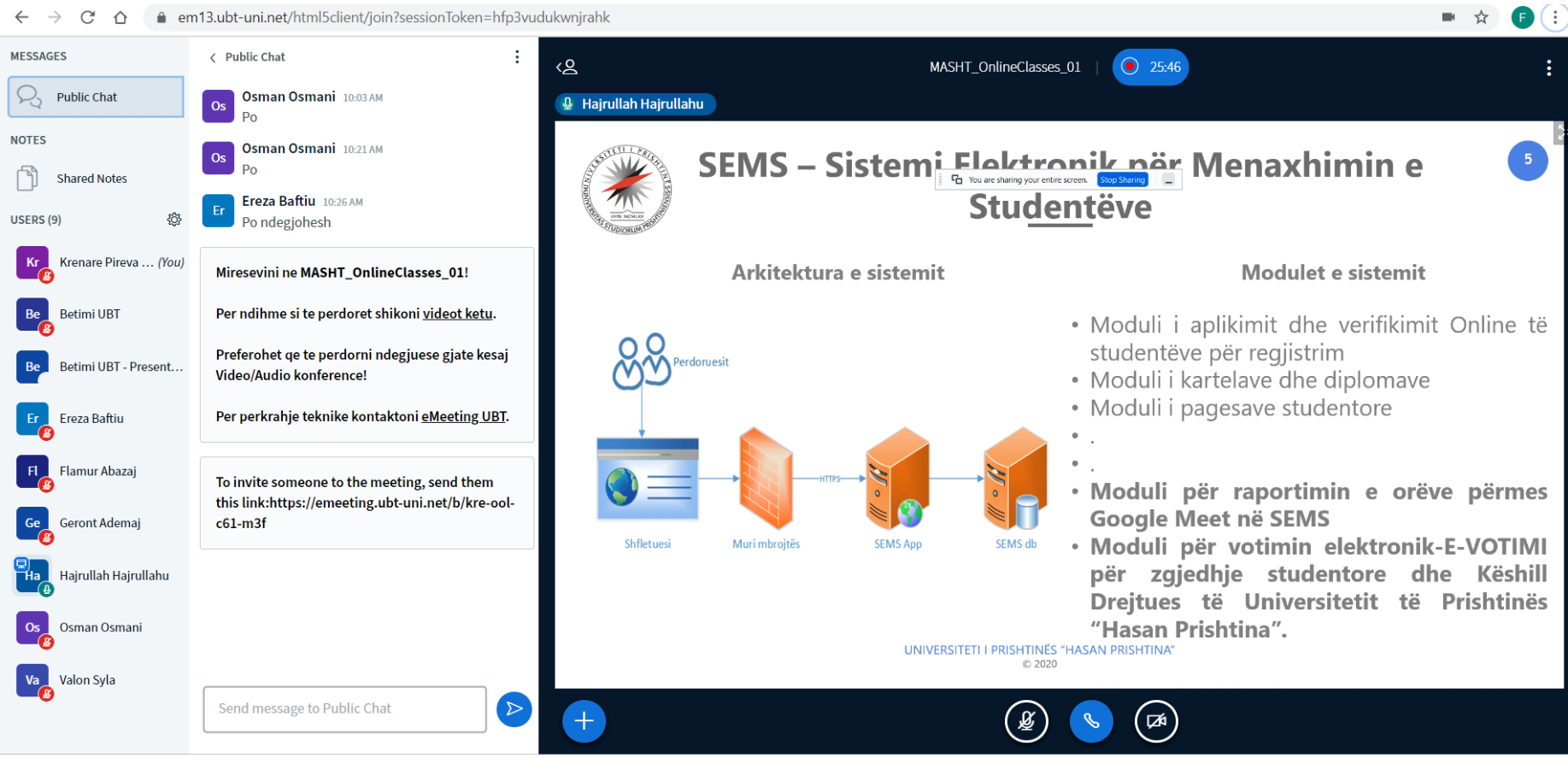}
\caption{A screenshot from the workshop} \label{Workshop}
\label{workshop}
\end{center}
\end{figure}

Firstly, to create an understanding of online teaching and learning we have gone through a number of papers described in the literature review.  There is defined a search strategy for collecting the most relevant papers by combining keywords for online learning, University infrastructure, video conferencing, online platforms for learning. 
Secondly, a workshop is organized on 22nd of December 2020 with main key stakeholders, respectively the chief technology officers of five Universities in Kosovo, with the aim to address the readiness of the technology infrastructure of Universities within Kosovo to provide online teaching during the COVID-19 pandemic. The workshop is organized online due to COVID situation using the Big Blue Button and a screenshot is presented in (Figure \ref{Workshop})

This workshop is supported by a project funded by the Ministry of Education Science and Technology in Kosovo and it took the initiative to invite chef technology officers (CTO) from:

\begin{enumerate}
    \item University for Business and Technology 
    \item University of Pristina
    \item University of Prizren
    \item University of Gjilan
    \item IBCM Mitrovica
\end{enumerate}

The discussion has been recorded, and later on, the transcript was documented.  For each CTO was allocated 15 minutes presentation which described the current University infrastructure, the platforms used for online learning, the challenges they experienced while supporting the switching process from in-campus classes into online classes. The workshop concluded with a one-hour discussion by tending to exchange the experiences among the participants. At later stages, the author conducted 3 interviews with a key expert of IT and discussed our findings from the collected data, which supported validating the findings. 

\subsection{Data Analysis}
\begin{figure}
\begin{center}
\includegraphics[width=150mm]{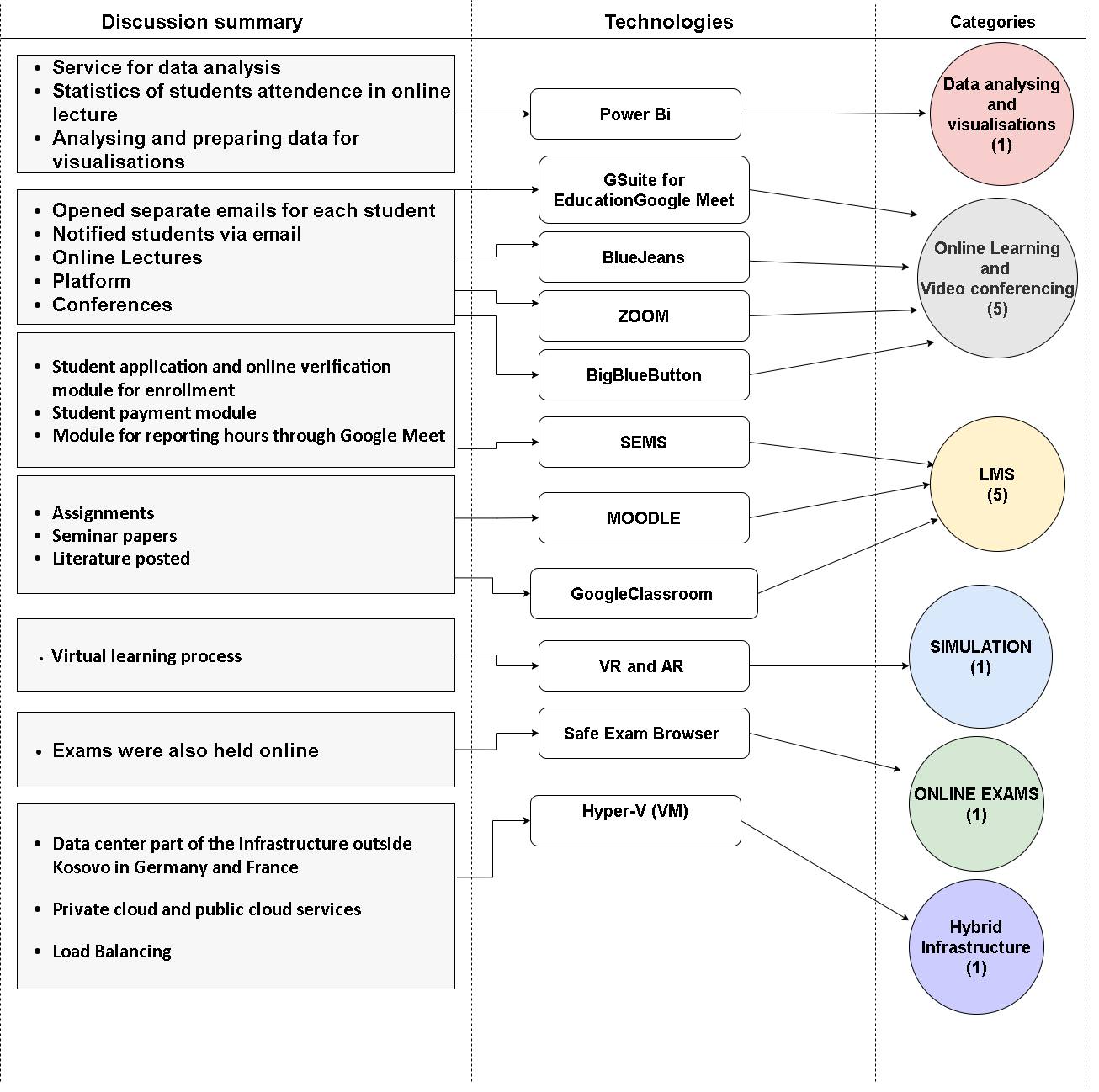}
\caption{ Qualitative analysis of experts discussion} 
\label{Litchman3C}
\end{center}
\end{figure}

Having in mind that for data collection is used a qualitative method, during this research we reviewed a number of data analysis methods and came up with the Litchmann 3C approach \cite{Litchman3C}, which is similar to the approach followed in \cite{krenarepireva}. The documented transcript is text-based and using the Litchmann 3C approach \cite{Litchman3C} we resulted in six categories. 
Figure \ref{Litchman3C}. presents the results of data analysis in three columns. The first column lists the summary of the most frequent discussion within the workshop, the second column lists the technologies derived from the summary of the discussion in column 1, and the third column extracts the categories that the participants mostly discussed.

Categories such as the Power Bi platform, the platform used the University of Pristina to analyze and to prepare data for visualization. Using this platform they also generated statistics of students’ attendance in an online lectures.
Each of the five participants’ universities has been using different platforms in order for online learning and video conferencing. Such platforms as G Suite for Education Google Meet used by all the universities. The Blue Jeans platform used only by IBCM College in Mitrovica, whereas the Big Blue Button platform is used by University for Business and Technology. And finally, IBCM Mitrovica claimed to have used the Zoom platform for online classes. All these platforms generated the Online learning and Video Conferencing category (See Figure \ref{Litchman3C}).
Learning Management System (LMS) category is created from three different platforms that universities declared that they are using in order to provide the learning materials, such platforms as SEMS (Sistemi Elektronik për Menaxhimin e Studenteve), Moodle, and Google Classroom. SEMS is developed by the University of Pristina, and they shared it with the University of Prizren and the University of Gjilan. Moodle was used by all of the universities to publish learning materials for students online, for assignments and seminar papers, whereas IBCM used the Google Classroom platform.
In addition, UBT College claimed that they have used additionally the virtual and augmented reality for simulation during the online classes. The same platforms are used also for the online exams, except that UBT integrated also the Safe Exam Browser for avoiding potential misuse of the exam time period.

\section{Findings and Discussion}
\label{sec:results}
In this section the findings are presented with respect to the key categories that resulted from the data analysis explained in the previous section.
\subsection{Before COVID-19 Phase}
Since 2019 UBT made an agreement with EON Reality and they started to use the virtual and augmented reality services prior to the pandemic. They have also some experiences before the pandemic for online learning, since there have been cases where students have organized online learning, however this situation was completely different since UBT took the decision to switch to online classes 100 percent due to the COVID-19 lockdown situation. Since 2010, the Moodle platform has been implemented and professors and students contributed with the learning materials, submission of seminars and assignments. After the implementation of Moodle, they extended their services with G suite for Education, by providing email account for all professors and students. Further, UBT developed various systems one of which is the Student Management Information System (SMIS) for additional administrative services. UBT has started researching various online platforms in the past in order to create the system for online training and certifications. Same approach was followed also by University of Pristina, who created the SEMS system for administrative services, and they used G Suits for Education as well. The SEMS systems were offered also to the rest of public Universities, to University of Prizren (Ukshin Hoti University) and University of Gjilan (University Kadri Zeka).
University of Pristina claimed that they experienced to have online classes prior to COVID situation but very rare, and the same was also claimed by the University of Prizren, which emphasized some classes with Croatia and Bosnia Universities. However, they claimed that in two years, an average of six lectures per semester have been organized online, and due to this they created two conferencing rooms just for online learning.	University of Gjilan, developed SMU for administrative issues and Moodle for sharing the learning materials, whereas the International Business College Mitrovica (IBCM) before the pandemic was mainly focused on the traditional setting where they held lectures physically together with students and professors.

As we can conclude from the previous discussion the situation of the pandemic caught the Universities in Kosovo almost unprepared for this situation when considering the process of  switching the in campus classes into online classes totally, 100\%. 

\subsection{During COVID-19 Phase}
\subsubsection{Data Analyzing and Visualisation}

The University of Pristina has used Power BI Desktop for analyzing and preparing data for visualization, where Power BI Desktop is connected to Power BI Service which prepared the Dashboard and published the data. All together related to various media such as: email, website, smartphone, tablet, Laptop / PC, TV. Power BI Service has displayed all statistics files in SEMS, where SEMS has displayed the data to the user, and then sent the data to the SEMS database. UBT College, student’s attendance in lectures and exercises was reported through the BigBlueButton platform, as for other universities, for the exact purpose of data analyzing and visualizations of students attendance they used the features of the video conferencing platforms.

\subsubsection{Video Conferencing}

Online learning at UBT was developed with the BigBlueButton platform which is an open source platform and was integrated within Moodle LMS during quarantine. UBT within next two weeks of March started online learning using BigBlueButton, initialy by testing the platform internally among  the academic staff and the students. After the first week, UBT started to organize the online classes along with exercises for the students. The manuals and videos of "how to use the BigBlueButton platform" was sent to staff and students prior to the first online classes. Students have access to the BigBlueButton platform only through official UBT emails therefore students that aren’t part of UBT do not have access to online lectures. This method has helped maintain the integrity of online learning. Student’s attendance in lectures and exercises was reported through the same platform. 

University of Pristina, University of Prizren and University of Gjilan during COVID-19 pandemic, used Google Meet platform for providing the online classes. University of Pristina also managed active users through Google Meet. During the pandemic over 2,000 teaching hours were held. Overall, they experienced to have 25,000 participants in online lectures where 1,500 of them were academic staff. Compared to other universities, it is worth mentioning that University of Gjilan implemented a strategy to allow students either with the official emails or unofficial emails to enter the online classes, with the intention to not miss the classes. 

IBCM college during the pandemics used BlueJeans platform for video conferencing and the reason for selecting BlueJeans has been its automation processes for providing the online learning education. Through BlueJeans they offered good opportunities to increase interactivity in lectures and in parallel with BlueJean they also used Zoom platform. The difficulty that IBCM has encountered during online lectures was problems with the internet and preventing direct access to students to help troubleshoot any possible problems. Later, since IBCM used G suits Education they provided the possibility to use Google meet as well. Through official college emails they have maintained the security of student access to the platforms knowing that the students who have accessed the lectures are IBCM students.

\subsubsection{Learning Management System (LMS)}
UBT faculty implemented Moodle, professors and students contributed in literature, materials, submission of seminars and assignments over this platform. They also used  the application of G suite for Education, where it has always been mandatory that every student who registers at UBT has the official email of UBT. Accessing the UBT system is done only through official email for security reasons so that students from different universities do not have the right to use and access. UBT has developed various systems one of which is the Student Management System (SMIS) for administrative services.

University of Pristina as well as University of Prizren and University of Gjilan, they used SEMS as a learning management system (LMS), which they created. 

The SEMS system during the pandemic evolved with a number of additional modules in order to comply with the pandemic rules, thus attempting to increase the number of services that the students can obtain online. The University of Pristina divided the system into several modules: 
\begin{itemize}
    \item Student application and online verification module for enrollment, 
\item Cards and diplomas module, 
\item Student payment module. 
\end{itemize}

During the pandemic they made the addition of modules: 
\begin{itemize}
\item	Module for reporting hours through Google Meet in SEMS, 
\item E-Voting module for student elections and Steering Council of the University of Pristina.
\end{itemize}

SEMS is used by professors and management in order to carry out all student administrative procedures in parallel. University of Gjilan also used Google Classroom and Moodle, so professors can publish literature for students and also students can submit their seminar papers and assignments through those platforms. The reporting for orginised lectures was done through Google Meet on SEMS. 

As for the IBCM college, they used as Learning Management System Google Suite for Education, Google Classroom. Through Google Classroom, students published their assignments.

\subsubsection{Simulation}
\begin{figure}
\begin{center}
\includegraphics[width=130mm]{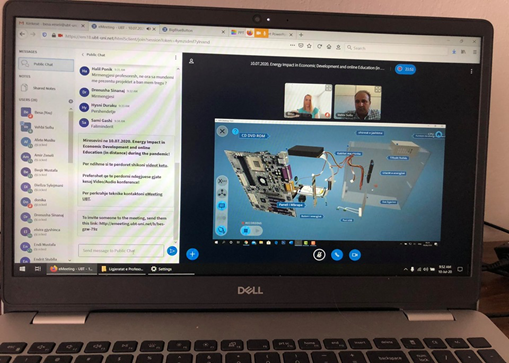}
\caption{Using AR and VR Platforms} \label{VR.png}
\end{center}
\end{figure}

During the pandemic UBT used virtual reality and augmented reality for simulation purposes. Online learning during the pandemic was also developed through ‘EON Reality’ platform in various faculties such as dentistry, nursing, mechatronics, to name a few (see Figure ~\ref{VR.png}).

\subsubsection{Online Exams}

After the online lectures, the period of exams was very challenging. In this respect, UBT implemented the Safe Exam Browser within Moodle which prevents the possibility of using other applications during the exam. Parallel to this, the students were requested to turn on their cameras through Google Meet, to make sure the students surrounding is clean until the exam is submitted. Prior to the exam, students have also received a manual report with all the necessary information and rules for holding the exams online in order to not have any misunderstanding. As for other universities of Kosovo, they held exams physically when the situation became more relaxed with the pandemic restrictions.

\subsection{Infrastructure}

\textbf{UBT} has developed its infrastructure based on good foundation and premise. In the Lipjan campus near capital city of Kosovo there is a server room where they have a number of servers located physically and a number of internet connection from various internet services providers due to inconsistency of the internet connection that we have here in Kosovo.  UBT claimed that part of their infrastructure is located outside of Kosovo, using a datacenter partially in Germany and France through Cloud Technology. This due to back-up data, as well as in case they are holding an online class and fall from one point to utilize the other point. UBT uses private cloud and public cloud services as well for a number of services that they provide to their staff and students. It all ended up in a hybrid infrastructure solution. Speaking in numbers, UBT has a total of eight servers. In terms of specification performance, 32 cores, 64 threads, 256 GB RAM with a sustainable power station. In addition to the main servers, UBT has used virtualization hosting services over 70 Virtual Machines(VM). Some of them are located inside Kosovo while some others are in a private cloud and public cloud.
During the pandemics, UBT experienced around 8,000 clicks concurrently, so they created a Clustered Database System, where data is located in several VM. Due to high number of concurrent users, UBT included load balancing, so they managed to balance the usage once experiencing high usage of services, concurrent wise. Even that the infrastructure supported simultaneously such high number of students, the session created for a single lecture limited the number of users, in our case the number of students for that particular session supported not more than 300.  
UBT implemented a strategy to record the lectures and provide the videos through Moodle, so the students can revisit the lectures in case they didn’t have the chance to be part of them. Therefore, currently UBT reached up to 2.5 TB of recorded lecture space, which are accessed by students only through the Moodle platform at any time. 

From the other side, the  \textbf{University of Pristina} (UP) used the services of G Suite for Education, where most of the information is stored in Google Cloud. For the development of online learning UP infrastructure consists of:
\begin{enumerate}
    \item twenty servers of the type Dell PowerEdge R410, 
  \item two servers of the type Dell PowerEdge R530, 
 \item six servers of the type Fujitsu Primergy RX2540 M1, 
 \item while the switch has two of the type CISCO SG200-25, 
 \item one switch of the type ZyXel The GS 1900-24, which has 1Gb /s bandwidth and 
 \item a D-Link Gigabit Switch DGS - 1024D, 
 \item two CICSO 2800 Series routers and one CISCO 3800 router.

\end{enumerate}

\begin{figure}
\begin{center}
\includegraphics[width=10cm]{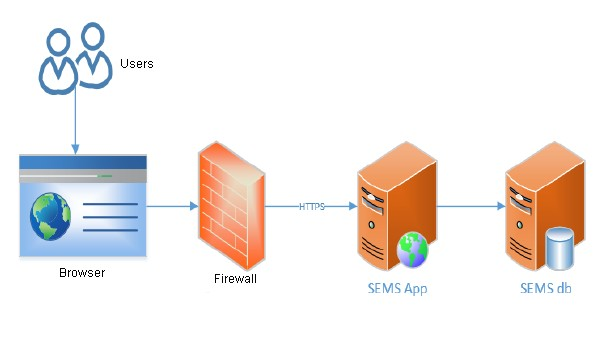}
\caption{System architecture of the University of Prishtina	} \label{UP.png}
\end{center}
\end{figure}

In (Figure ~\ref{UP.png}) is provided a simulation architecture that University of Pristina follows, in order to offer the services to the staff and students. So, the users (staff and students) accesses the browser, accesses the website, but in case of accessing sensitive modules it needs to pass through a firewall for filtering purposes, so only authorized users had the ability to access the system. 

In \textbf{University of Prizren} the situation is a bit different. The switching process had some difficulties due to the fact that the users experienced a lot of problems once approaching to access our systems. Their equipment lack of proper performance and lack of appropriate applications. Additionally, in Prizren region the internet service providers experienced a lot of downtime, interrupting the online classes continually. Their infrastructure is totally dependent in University of Prishtina, since we have inherited their infrastructure, and the additional services that we planned to offer to our distance students used Google for Education services. 

The infrastructure implemented in the  \textbf{University of Gjilan} contains two servers. One of them is located in the University of Pristina, so it is managed by UP. Meanwhile the other server is located in the University of Gjilan where Moodle is hosted. This infrastructure offers services for around 3,000 active students of the University of Gjilan and about 200 academic staff. For the rest of services they decided to go through the Google for Education.  To conclude, University of Gjilan using the current in campus infrastructure and Google Cloud Infrastructure organized 3,600 hours online per day, 50 lectures, by providing services for around 3,000 active students along with 97 teachers. During the ten-week period throughout the pandemic there were a total of 100,000 online connections and 1,500 daily connections.
Finally,  \textbf{IBCM} used G suite for education, along with BlueJeans and Zoom platforms, all as cloud services.

\section{Conclusion}
\label{sec:conslusion}
To sum up, everything that has been previously stated, this paper aimed to gather the right information about the transformation process from in-campus classes into online classes from the technical perspective. The research was supported by the Ministry of Education, Science, and Technology in Kosovo, with the intention to analyze how the Higher Education Institutions have been affected during the COVID-19 pandemic situation, with the main focus to their respective network infrastructure. Based on the results, the authors claim that there was no national plan on how to continue with online learning, thus, each University had its own plans when deciding to offer online learning. Most of the Universities used Google for Education services and additional platforms build to their internal LMS systems (example: BigBlueButton) and administrative system (example: SMIS, SEMS) in an agile manner, depending on the needs and requirements from the management, professors, and students as the main stakeholders in this situation. 

\bibliographystyle{splncs04}
\bibliography{main.bib}

\begin{thebibliography}{10}
\providecommand{\url}[1]{\texttt{#1}}
\providecommand{\urlprefix}{URL }
\providecommand{\doi}[1]{https://doi.org/#1}

\bibitem{erezabaftiuthesis}
The transformation process from in-campus classes into online classes due to
  covid-19 pandemic situation - the case of higher education institutions in
  kosovo, b.a.thesis. ubt higher education institution

\bibitem{adnan2020online}
Adnan, M., Anwar, K.: Online learning amid the covid-19 pandemic: Students'
  perspectives. Online Submission  \textbf{2}(1),  45--51 (2020)

\bibitem{ali2013network}
Ali, M.N.B., Rahman, M.L., Hossain, S.A.: Network architecture and security
  issues in campus networks. In: 2013 Fourth International Conference on
  Computing, Communications and Networking Technologies (ICCCNT). pp.~1--9.
  IEEE (2013)

\bibitem{arkorful2015role}
Arkorful, V., Abaidoo, N.: The role of e-learning, advantages and disadvantages
  of its adoption in higher education. International Journal of Instructional
  Technology and Distance Learning  \textbf{12}(1),  29--42 (2015)

\bibitem{bhakti2020web}
Bhakti, M.A.C., Wandy, W.: Web conference internet traffic analysis during
  study-from-home period: Case in sampoerna university. Indonesian Journal of
  Computing, Engineering and Design (IJoCED)  \textbf{2}(2),  91--98 (2020)

\bibitem{emrouznejad2010cooper}
Emrouznejad, A., De~Witte, K.: Cooper-framework: A unified process for
  non-parametric projects. European Journal of Operational Research
  \textbf{207}(3),  1573--1586 (2010)

\bibitem{kebritchi2017issues}
Kebritchi, M., Lipschuetz, A., Santiague, L.: Issues and challenges for
  teaching successful online courses in higher education: A literature review.
  Journal of Educational Technology Systems  \textbf{46}(1),  4--29 (2017)

\bibitem{kolar2020experiences}
Kolar, P., Tur{\v{c}}inovi{\'c}, F., Bojanjac, D.: Experiences with online
  education during the covid-19 pandemic--stricken semester. In: 2020
  International Symposium ELMAR. pp. 97--100. IEEE (2020)

\bibitem{krenarepireva}
K.P.Nuci, R.Tahir, A., A.S.Imran: Game-based digital quiz as a tool for
  improving sutdents engagement and learning in online lecture. Original
  Article,Paper submitted and is in progress  (2020)

\bibitem{krishnamurthy2020future}
Krishnamurthy, S.: The future of business education: A commentary in the shadow
  of the covid-19 pandemic. Journal of business research  \textbf{117}, ~1--5
  (2020)

\bibitem{kullberg2018implementing}
Kullberg, M., et~al.: Implementing remote customer service api using webrtc and
  jitsi sdk  (2018)

\bibitem{lichtman2013making}
Lichtman, M.: Making meaning from your data. Qualitative Research in Education:
  A User's Guide. 3rd ed. London: SAGE Publications, Inc p.~241 (2013)

\bibitem{meyer2010infrastructure}
Meyer, J.D., Barefield, A.C.: Infrastructure and administrative support for
  online programs. Online Journal of Distance Learning Administration
  \textbf{13}(3) (2010)

\bibitem{reimers2020framework}
Reimers, F.M., Schleicher, A.: A framework to guide an education response to
  the covid-19 pandemic of 2020. OECD. Retrieved April  \textbf{14}(2020),
  2020--04 (2020)

\bibitem{stefanov2020learning}
Stefanov, R.R., Ivanova, V.D., Grigorova, R.Z., Petkova, P.Y.: E-learning in
  the covid-19 context-epidemiological and educational challenges. In: 2020
  International Conference Automatics and Informatics (ICAI). pp.~1--6. IEEE
  (2020)

\end{thebibliography}

\end{document}